\documentclass{mn2e}
\usepackage{graphicx}
\usepackage{amsbsy}
\usepackage{amsmath}
\usepackage{epsf}
\usepackage{paralist}
\usepackage{tablefootnote}
\usepackage{hyperref}
\usepackage{tabularx}
\usepackage{kantlipsum}
\usepackage{upgreek}

\newcommand{\apj}{ApJ}
\newcommand{\apjl}{ApJL}
\newcommand{\mnras}{MNRAS}
\newcommand{\apjs}{ApJS}

\newcommand{\aap}{A\&A}

\newcommand{\solphys}{SoPh}
\newcommand{\araa}{Annu. Rev. Astron. Astrophys.}

\newcommand{\Alf}{Alfv\'{e}n}

\newcommand{\beq}{\begin{equation}}
\newcommand{\eeq}{\end{equation}}

\DeclareMathAlphabet{\mathsfsl}{OT1}{cmss}{bx}{sl}
\SetMathAlphabet{\mathsfsl}{bold}{OT1}{cmss}{bx}{sl}

\usepackage{amssymb}
\usepackage{pifont}

\usepackage{epstopdf}
\usepackage{color}        

\begin{document}

\title[Physical properties of clouds from striations]
{A new method for probing magnetic field strengths from striations in the interstellar medium}

\author[Tritsis et al.]
  {Aris Tritsis$^{1}$, Christoph Federrath$^{1}$, Nicola Schneider$^{2,3}$, Konstantinos Tassis$^{4,5}$ \\
    $^1$Research School of Astronomy and Astrophysics, Australian National University, Canberra, ACT 2611, Australia\\
    $^2$I. Physik. Institut, University of Cologne, Z\"{u}lpicher Str. 77, 50937 Cologne, Germany\\
    $^3$OASU/LAB, Universit\'e de Bordeaux, 33615 Pessac, France\\
    $^4$Department of Physics and ITCP, University of Crete, PO Box 2208, 71003 Heraklion, Crete, Greece\\
    $^5$IESL and Institute of Astrophysics, Foundation for Research and Technology-Hellas, PO Box 1527, 71110 Heraklion, Crete, Greece\\}
\maketitle 

\begin{abstract}

Recent studies of the diffuse parts of molecular clouds have revealed the presence of parallel, ordered low-density filaments termed striations. Flows along magnetic field lines, Kelvin-Helmholtz instabilities and hydromagnetic waves are amongst the various formation mechanisms proposed. Through a synergy of observational, numerical and theoretical analysis, previous studies singled out the hydromagnetic waves model as the only one that can account for the observed properties of striations. Based on the predictions of that model, we develop here a method for measuring the temporal evolution of striations through a combination of molecular and dust continuum observations. Our method allows us to not only probe temporal variations in molecular clouds but also estimate the strength of both the ordered and fluctuating components of the magnetic field projected on the plane-of-the-sky. We benchmark our new method against chemical and radiative transfer effects through two-dimensional magnetohydrodynamic simulations coupled with non-equilibrium chemical modelling and non-local thermodynamic equilibrium line radiative transfer. We find good agreement between theoretical predictions, simulations and observations of striations in the Taurus molecular cloud. We find a value of $\rm{27 \pm 7} ~\rm{\upmu G}$ for the plane-of-sky magnetic field, in agreement with previous estimates via the Davis-Chandrasekhar-Fermi method, and a ratio of fluctuating to ordered component of the magnetic field of $\sim$ 10\%. 

\end{abstract}


\section{Introduction}\label{intro}

Molecular clouds are usually observed to be complex, turbulent systems (Heyer \& Brunt 2004; Heyer et al. 2006; Falgarone et al. 2009; Andr{\'e} et al. 2014) with internal velocities ranging from $\sim$$\rm{0.2}$~km/s in prestellar cores (Goodman et al. 1998), up to a few km/s on their largest scales (Larson 1981; Solomon et al. 1987; Ossenkopf \& Mac Low 2002; Roman-Duval et al. 2011). Given their internal velocities, observing temporal variations of even the smallest resolved structures through a sequence of observations is prohibitive for one lifetime. Robust measurements of the temporal variations in molecular clouds could provide invaluable information in determining their lifetimes and the timescales of star formation (Tassis \& Mouschovias 2004). 

In stark contrast to the general turbulent nature of molecular clouds (Mac Low \& Klessen 2004; McKee \& Ostriker 2007; Padoan et al. 2014; Federrath \& Klessen 2012), a recently discovered type of low-density, elongated structures is always observed to be well-ordered, quasi-periodic and well-aligned with the plane-of-the-sky (POS) magnetic field (Goldsmith et al. 2008; Miville-Desch{\^e}nes et al. 2010; Palmeirim et al. 2013; Alves de Oliveira et al. 2014; Cox et al. 2016; Malinen et al. 2016; Panopoulou et al. 2016). Because of their highly ordered nature, these structures, termed striations, \textit{are unlikely to form from a turbulent process}. Striations are a common feature amongst interstellar clouds and are often associated with denser filaments (Palmeirim et al. 2013; Alves de Oliveira et al. 2014; Cox et al. 2016; Malinen et al. 2016), and the sites of star formation. 

Striations were first seen in the Taurus molecular cloud (Narayanan et al. 2008; Goldsmith et al. 2008) and since their discovery, a number of physical mechanisms were proposed to explain their formation. The most popular explanation was that they are created because of flows along magnetic field lines (Miville-Desch{\^e}nes et al. 2010; Palmeirim et al. 2013; Alves de Oliveira et al. 2014; Neha et al. 2018). Other possibilities included a Kelvin-Helmholtz instability either along or perpendicular to magnetic field lines (Heyer et al. 2016; Tritsis \& Tassis 2016) and hydromagnetic waves (Goldsmith et al. 2008; Heyer et al. 2016; Tritsis \& Tassis 2016). For an alternative formation mechanism see Chen et al. (2017).

Tritsis \& Tassis (2016) performed a combined theoretical, numerical and observational analysis of striations considering four different physical models including sub-\Alf ic and super-\Alf ic flows along field lines, a Kelvin-Helmholtz instability, and that of hydromagnetic waves. They found that the models involving flows along field lines and that of a Kelvin-Helmholtz instability produced striations-like structures but with such an extremely small contrast between adjacent striations that they would not be observable. Specifically, the maximum contrast in these models was $\sim$0.01\% compared to $\sim$25\% in the observations. Additionally, they found that the power spectra of cuts perpendicular to the striations-like structures formed were \textit{not} in agreement with observations. 

Tritsis \& Tassis (2016) also examined a model in which striations are formed as a result of fast magnetosonic waves compressing the gas and creating ordered structures parallel to magnetic field lines. In this model, fast magnetosonic waves are excited as a result of phase mixing with \Alf ~waves (Nakariakov et al. 1997). When second-order terms in the linearized MHD equations can be neglected, \Alf ~waves and fast magnetosonic waves (if at all present in the system) propagate independently. However, when second-order terms are non-negligible, the equation governing the nonlinear dynamics of fast magnetosonic waves (Nakariakov \& Oraevsky 1995) will contain terms associated with the perturbed quantities of the \Alf ~wave. Phase mixing of \Alf ~waves with compressible modes has been extensively studied in the context of solar physics and the heating of the solar corona (e.g. Heyvaerts \& Priest 1983; Lee \& Roberts 1986; Parker 1991; Nakariakov et al. 1997) and magnetospheric physics (e.g. Malara et al. 1996) both numerically and analytically and in the presence of density inhomogeneities has been found to be strong (however, see also Cho \& Lazarian 2002). Based on numerical simulations, Tritsis \& Tassis (2016) found that the maximum contrast between adjacent striations for this model to be comparable to observations. They also found that the power spectra of cuts perpendicular to striations were in agreement with observations. They concluded that out of the four models the only mechanism consistent both qualitatively and quantitatively with observations of striations is the one that involves the excitation of fast magnetosonic waves.

From the physical model developed by Tritsis \& Tassis (2016) a number of predictions arose. These predictions can be tested observationally. One prediction is that, in the presence of boundaries, these hydromagnetic waves can get trapped, setting up normal modes. This prediction was recently confirmed in the case of the molecular cloud Musca (Tritsis \& Tassis 2018). The discovery of normal modes is confirmation of the hydromagnetic-waves model, since this scenario cannot be realized under some turbulent process.

Apart from striations, magnetic fields influence the evolution and formation of the densest parts of molecular clouds. However, the exact role of magnetic fields and their relative importance compared to other processes is still under debate (e.g. Federrath 2016a). Robust measurements of the field strength, free from instrumental effects, are essential in order to set constraints on theoretical models and simulations.

The two most popular techniques for measuring magnetic fields in molecular clouds are through polarization measurements and the Davis, Chandrasekhar \& Fermi (DCF) method or variations of it (Davis 1951; Chandrasekhar \& Fermi 1953; Hildebrand et al. 2009), and the Zeeman effect (for a review on magnetic field measurements in the ISM see Crutcher 2012). However, in both of these methods, instrumental effects can significantly affect the final results. Zeeman observations require long integration times in order to achieve the high signal-to-noise ratios (SNR) necessary, and a gain difference between left and right circular polarizations can introduce a signal even if it is not present. In the case of polarimetric observations, an intrinsic spread of polarization angles due to systematics can lead to an underestimation of magnetic field strengths. Recently, Gonz{\'a}lez-Casanova \& Lazarian (2017) developed a new method for studying magnetic fields through velocity gradients from spectroscopic observations. However, their method currently does not take into account optical depth effects. 

In the present paper, we test a second prediction from the physical model developed by Tritsis \& Tassis (2016) (see \S~\ref{method}) and develop a method for measuring the time-scales of striations and the ordered and fluctuating components of the magnetic field. In \S~\ref{method} we provide the theoretical background and describe the method. Numerical simulations and synthetic observations used to benchmark our analysis are presented in \S~\ref{sims}. In \S~\ref{observations} we employ $Herschel$ dust emission maps to create a new column density map of the region with striations in Taurus. We use this map and $^{12}$CO (J = 1 -- 0) line emission data to derive the time-scales of the striations in Taurus and estimate the magnetic field. We discuss our results in \S~\ref{discuss} and summarize in \S~\ref{sum}.


\section{Method}\label{method}
Since striations are created by fast magnetosonic waves, measurements of their time-scales, namely the angular frequencies of the waves, can be used to measure the magnetic field strength through their dispersion relation:
\begin{equation}\label{disprel}
\frac{\omega}{k} = \sqrt{v_A^2 + c_s^2}
\end{equation}
In Equation~\ref{disprel}, $\omega$ is the frequency of the waves, $k$ is the wavenumber, $c_s$ is the sound speed and $v_A$ is the \Alf~speed. Here, we ignored waves propagating in directions other than exactly perpendicular to the magnetic field (see \S~\ref{discuss} for a discussion on our assumptions). By substituting the expression for the \Alf~speed ($v_A = B_0/\sqrt{4\pi\rho_0}$, where $B_0$ is the mean magnetic field and $\rho_0$ is the mean density), ignoring the sound speed ($c_s^2\ll v_A^2$) (see \S~\ref{discuss}) and rearranging we get:
\begin{equation}\label{disprel2}
B_0 = \frac{\omega}{k}\sqrt{4\pi\rho_0}
\end{equation}

In regions of striations, the displacement of the gas at every position $\textbf{r}$ and time $t$ will be given as a superposition of waves:
\begin{equation}\label{displ}
\mathbf{\rm{\xi}}(\mathbf{r}, t) = \sum_n A_n e^{i(\mathbf{k_nr}-\omega_n t)}
\end{equation}
where $A$ is the amplitude and the subscript $n$ denotes different wave modes. Taking the derivative of the displacement we get:
\begin{equation}\label{vel}
\mathbf{v_1}(\mathbf{r}, t) = -\sum_n i A_n\omega_n e^{i(\mathbf{k_nr}-\omega_n t)}
\end{equation}
where $v_1$ is the perturbed velocity of the region and we carry out our analysis in the frame of reference of the cloud such that the bulk velocity ($v_0$) is zero. Furthermore, by linearising the continuity equation, ignoring second-order terms and integrating we find (Spruit 2013):
\begin{equation}\label{cont}
\rho_1 = \rho_0\mathbf{\nabla}\cdot\mathbf{\rm{\xi}}(\mathbf{r}, t)
\end{equation}
where $\rho_1$ is the perturbed density. By performing the same analysis for the induction equation of ideal magnetohydrodynamics (MHD) and assuming that the velocity along field lines in such regions is much smaller than the perpendicular velocity components (see \S~\ref{discuss}) we obtain:
\begin{equation}\label{induct}
B_1 = B_0\mathbf{\nabla}\cdot\mathbf{\rm{\xi}}(\mathbf{r}, t)
\end{equation}
where $B_1$ is the magnitude of the fluctuating component of the magnetic field. Combining Equations~\ref{cont} \&~\ref{induct} we get:
\begin{equation}\label{ratio}
\frac{B_1}{B_0} = \frac{\rho_1}{\rho_0} = \frac{N^1_{\mathrm{H_{2}}}}{N^0_{\mathrm{H_{2}}}}
\end{equation}
where $N\rm{^1_{H_{2}}}$ and $N\rm{^0_{H_{2}}}$ are the fluctuating and mean column density, respectively. Consequently, the ratio of the fluctuating to the mean magnetic field in the striations regions, can be computed from the respective ratio of column densities. From Equations~\ref{displ} \& \ref{cont} and the fact that $N\rm{^1_{H_2}} = \int_0^{L_{los}} \rho_1 dl$, where $L_{\rm{los}}$ is the line-of-sight (LOS) dimension,  we find:
\begin{equation}\label{cd}
N^1_{H_2} = N^0_{H_2}\sum_n A_n k_n ie^{i(\mathbf{k_nr}-\omega_n t)}\,.
\end{equation}

From this analysis it becomes apparent that the power spectra of velocity and the power spectra of column density cuts perpendicular to the striations should peak at the same positions, since the same wavemodes $k_n$ appear in Equations~\ref{vel} \&~\ref{cd}. The wavenumbers can be easily computed by considering the power spectrum of column density cuts perpendicular to the long axis of the striations. The power in the power spectrum of such cuts would be $\vert P_{N^1_{H_{2}}} \vert^2=$  $\vert N^0_{H_2} A_n k_n\vert^2$ where $\vert P_{N^1_{H_{2}}} \vert^2$ is the energy spectral density. Similarly, the power in the power spectrum of the velocity $\vert P_{v_1} \vert^2$ should be $\vert\omega_n A_n\vert^2$. We define the parameter $\Gamma$ as the square root of the ratio of the two powers:
\begin{equation}\label{definition}
\Gamma_n = \sqrt{\frac{\vert P_{v_1} \vert^2}{\vert P_{N^1_{H_{2}}} \vert^2}} =  \frac{\omega_n}{k_n N^0_{H_{2}}}\,.
\end{equation}
The velocity power spectrum can be obtained from velocity centroids from spectroscopic observations. However, since we can only observe the LOS velocity component, when the magnetic field is at an angle with respect to the plane-of-the-sky (POS), the observed velocity will be smaller by $\rm{cos\uptheta}$ where $\uptheta$ is the angle between the magnetic field direction and the POS. Thus, taking projection effects into account and combining Equation~\ref{definition} and Equation~\ref{disprel2} we get that for any mode $n$:
\begin{equation}\label{mainEq}
B_0\mathrm{cos}\theta = B^{pos}_0 = \Gamma_n N^0_{H_{2}} \sqrt{4\pi \rho_0}
\end{equation}
where all quantities on the right-hand side of Equation~\ref{mainEq} are observables. The mean column density $N\rm{^0_{H_{2}}}$ can be computed from column density maps derived from dust emission observations; the parameter $\Gamma$ from the power spectra of column density cuts and velocity centroid cuts perpendicular to the striations and estimates for the mean density can be computed from multi-level observations of an individual molecule. In what follows, we use Equation~\ref{mainEq} to derive the magnetic field strength from synthetic and actual observations.


\section{Simulations}\label{sims}

\subsection{Simulation code}

In order to validate and test our method against chemical and radiative transfer effect, we perform numerical simulations using the astrophysical code FLASH 4.4 (Fryxell et al. 2000; Dubey et al. 2008). The equations of ideal MHD without gravity and with an isothermal equation of state are solved on a 2D cartesian grid using the unsplit staggered mesh algorithm (Lee et al. 2009; Lee 2013). For accuracy, we use Roe's solver (Roe 1981) for the Riemann problem. We use van Leer's flux limiter (van Leer 1979) and third-order interpolation. We adopt open boundary conditions (``diode") in order to minimize any effects from boundaries. Simulations are performed on a 256$\times$256 uniform grid and an additional simulation with half that resolution is performed to ensure convergence (see Appendix~\ref{convergence}). 

\subsection{Initial conditions}

\begin{figure}
\includegraphics[width=1.0\columnwidth, clip]{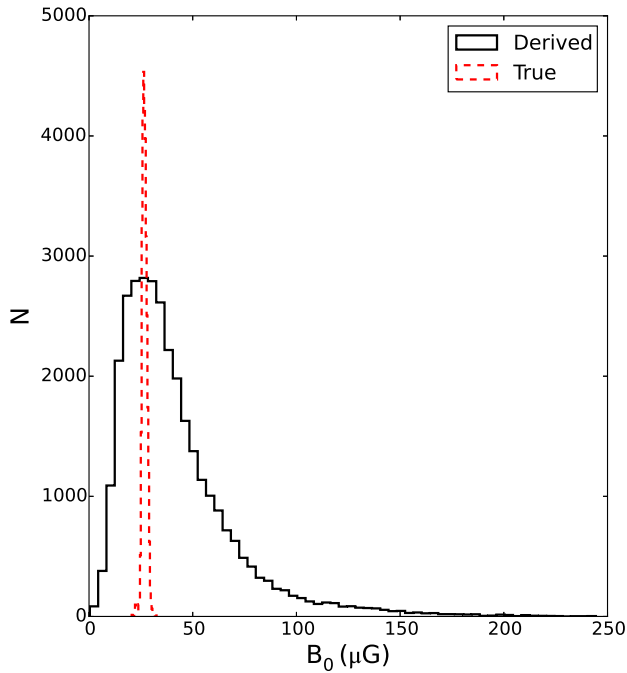}
\caption{Magnetic field values derived from our numerical simulations using Equation~\ref{simpler2} (black histogram) and the true magnetic field values directly extracted from the simulations (dashed red histogram). The two distributions peak at the same magnetic field value.
\label{SimplerDist}}
\end{figure}

\begin{figure}
\includegraphics[width=1.0\columnwidth, clip]{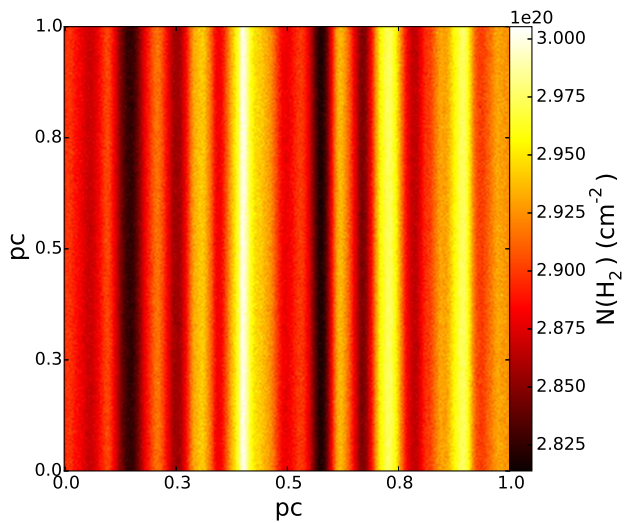}
\caption{Synthetic column density map of striations from our numerical simulations (noise is added in the density field with SNR = 50).
\label{striationsCDSims}}
\end{figure}

The size of our computational region ($\textit{L}$) is set equal to 1 pc in the $x$ and $y$ directions. The initial density is set to 100 $\rm{cm^{-3}}$ and the temperature is 15 K. The magnetic field is oriented perpendicular to this sheet, in the $z$ direction. In our reference run the value of the magnetic field is set to 30 $\rm{\upmu G}$. The magnetic field value, the temperature and density are selected such that they represent typical values of molecular cloud striations from observations (Goldsmith et al. 2008; Chapman et al. 2011; Heyer et al. 2016; Panopoulou et al. 2016). We additionally run two more simulations with magnetic field values of 10 and 100 $\rm{\upmu G}$ keeping the rest of the parameters constant. In our initial conditions we introduce an oscillating magnetic field given by:
\begin{equation}\label{Bsetup}
B_1 = B_0\sum_{n=1}^6A_n\mathrm{cos}(k_nr)
\end{equation}
where the wavenumbers are drawn from a uniform distribution in the range [$\rm{2\uppi/\textit{L}}$, $\rm{16\uppi/\textit{L}}$] and $r = \sqrt{x^2+y^2}$. The amplitude $A_n$ of each wavemode is linearly decreasing as a function of wavenumber with the total strength of the perturbed magnetic field limited to 30\% of $\rm{B_0}$. The amplitudes of the wavemodes are not selected based on some physical reason. Instead, their selection is simply a choice for the particular simulations performed here. In fact, power spectra of striations from the Musca molecular cloud suggest that some small scales have more power than the larger ones, although this may be an effect of the phase of each wavemode (Tritsis \& Tassis 2018). However, our results are independent of this choice (see \S~\ref{method}) and in general, larger scales tend to have more power than smaller ones (Tritsis \& Tassis 2016; Tritsis \& Tassis 2018). The initial density field setup is such that the plasma $\upbeta$ (defined as $\upbeta=8\pi P_{\rm{th}}/B^2$ where $P_{th}$ is the thermal pressure) is constant everywhere. For our reference run, the plasma $\upbeta$ is $\sim$ $1.4\times10^{-2}$, for our 100 $\rm{\upmu G}$ simulation is $1.3\times10^{-3}$ and for our 10 $\rm{\upmu G}$ simulation is $1.3\times10^{-1}$.

The angular frequency for each wave mode is then computed from the dispersion relation (Equation~\ref{disprel}) and the perturbed velocity was initially set in a self-consistent manner by:
\begin{equation}\label{Vsetup}
v_1 = \sum_{n=1}^6A_n\omega_n\mathrm{cos}(k_nr)
\end{equation}

\subsection{Chemical evolution}

Our dynamical simulations are also coupled with non-equilibrium chemistry. The chemical network used, including a table with all species considered and the initial elemental abundances, is described in detail in Tritsis et al. (2016). Here, we give a brief summary. In each time-step $\sim$ 14000 chemical reactions are solved and the abundances of $\sim$ 300 species are updated in each cell of our computational domain. Reaction rates for our chemical network are taken from the \textsc{UMIST} database (McElroy et al. 2013). The mean molecular weight is $\sim$ 2.4, the cosmic-ray ionization rate is set equal to $\rm{\zeta} =  1.3\times10^{-17} ~s^{-1}$ and the visual extinction for the simulations described here is set to $\rm{A_v}$ = 1 mag.

\begin{figure*}
\includegraphics[width=2.2\columnwidth, clip]{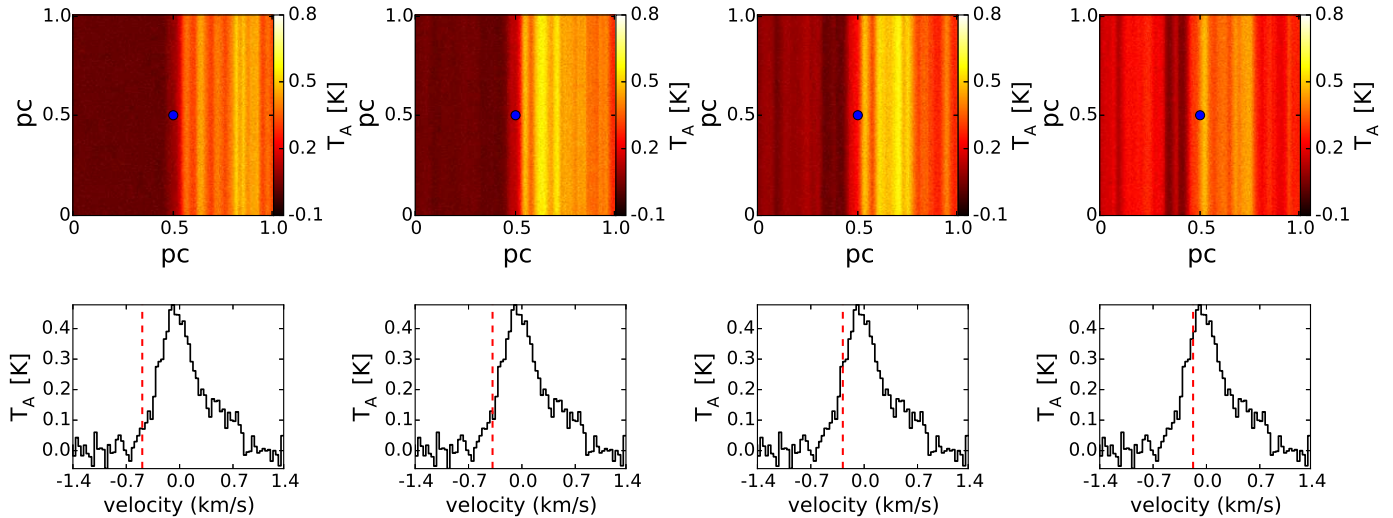}
\caption{$^{12}$CO (J = 1 -- 0) synthetic observations of striations from our astrochemical and non-LTE radiative transfer simulations. The upper panel shows four velocity slices and in the lower panel we plot a spectrum (black line) threading the center of our computational region (marked with blue points in the upper row). The red dashed lines overplotted on the spectrum mark the velocity of the velocity slices shown in the upper panel. The velocity range is set by the number of velocity bins and the spectral resolution. 
\label{vslices}}
\end{figure*}

\subsection{Simulation Results}

By assuming plane-wave solutions for all quantities and substituting in the continuity equation we find:
\begin{equation}\label{simpler1}
\frac{\omega}{k}\rho_1 = \rho_0 v_1
\end{equation}
Ignoring again the sound speed ($c_s \ll v_A$), substituting $\omega/k$ from the dispersion relation (Equation~\ref{disprel}), the definition of the \Alf ~speed and rearranging we find:
\begin{equation}\label{simpler2}
B_0 = \frac{\sqrt{4\pi\rho_0^3}v_1}{\rho_1}
\end{equation}
which can be written in terms of the fluctuating and mean column density. 

Using Equation~\ref{simpler2} we derive the magnetic field value from our simulations. The fluctuating density $\rho_1$ is computed as the density of each grid point minus the mean density inside our computational region. However, for grid points with density equal to the mean density in the computational region the derived magnetic field value will be infinite. To avoid this problem, we only consider grid points for which the absolute difference of the local density to mean density is equal or higher to 5 $\rm{cm^{-3}}$. Due to the open boundary conditions, the magnetic field will diffuse and its value will be slightly less than the initial setup.

In Figure~\ref{SimplerDist} we show the histogram of magnetic field values derived from Equation~\ref{simpler2} (black line) and the true values of the magnetic field (dashed red histogram). For visualization purposes, the binwidth in the derived histogram of magnetic field values is 10 times larger than in the true one. The true mean value of the magnetic field (red distribution) is 27 $\rm{\upmu G}$ and the standard deviation is 1 $\rm{\upmu G}$. The mode of derived (based on Equation~\ref{simpler2}) magnetic field values (black histogram), along with the 16th and 84th percentiles, is $26^{+37}_{-7}$ $\rm{\upmu G}$. Following the same procedure, the derived magnetic field value in our 100 $\rm{\upmu G}$ simulation is $61^{+110}_{-21}$ $\rm{\upmu G}$ compared to the true value of $79\pm4$ $\rm{\upmu G}$. Similarly, for our 10 $\rm{\upmu G}$ simulation the derived magnetic field value is $10^{+12}_{-4}$ $\rm{\upmu G}$ compared to the true value $9.3\pm0.6$ $\rm{\upmu G}$. Thus, although the errors using this simple method are large, the most probable value of the magnetic field is reasonably well recovered. 

\section{Application of the method to synthetic observations}

\subsection{Producing synthetic observations}

From our 2D simulations we produce 3D density cubes by duplicating the sheet multiple times along the direction of the magnetic field until the third dimension has equal size as the other two. A column density map of striations from our numerical simulations is shown in Figure~\ref{striationsCDSims}. Gaussian noise with signal-to-noise ratio of 50 is added to the density field. The contrast between adjacent striations in column density (1 -- 2 \%) is low compared to column density maps of striations in observations (5 -- 10 \%) (Palmeirim et al. 2013; Cox et al. 2016; see also Figure~\ref{striationsCD}). This is mainly because of the open boundary conditions, which, given enough time, will result in uniform density and magnetic field values in the computational region. The initial amplitude of the perturbations introduced in the system also affects the final contrast. Thus, if the strength of the perturbed magnetic field was more than 30\%, the contrast would be higher. However, for these numerical experiments, it is only the ratio of contrasts in column density and velocity that is important.

\subsection{Non-LTE line transfer to generate synthetic spectra}

We post-process the 3D density cubes produced by our dynamical simulations using the multi-level, non-local thermodynamic equilibrium (non-LTE) line radiative transfer code \textsc{PyRaTE} (Tritsis et al. 2018). \textsc{PyRaTE} takes as input the density, the temperature, the velocity field and molecular abundance. Velocities along the direction of the magnetic field are set to zero. For computing the population densities \textsc{PyRaTE} uses the escape probability approximation. For a cartesian grid, the algorithm searches the minimum optical depth towards 6 directions ($\pm$ x, $\pm$ y, $\pm$ z) by summing the infinitesimal optical depths of all grid points for which their absolute velocity difference is smaller than the thermal width. We produce position-position-velocity (PPV) cubes assuming non-LTE conditions for our reference run where the initial magnetic field value is $\rm{30~ \upmu G}$ and LTE conditions for the $\rm{10}$, $\rm{100~ \upmu G}$ and convergence test runs. 

\subsection{Results}

Four velocity slices from our synthetic observations of a $^{12}$CO (J = 1 -- 0) Position-Position-Velocity (PPV) cube are shown in the upper panel of Figure~\ref{vslices}. In the lower panel we show a spectrum from a LOS passing through the center of the emission map in each velocity slice and a line tracing the velocity. Results are plotted in antenna temperature units. Radiative transfer calculations are performed assuming a spectral resolution of 0.035 km/s, but the cube is then smoothed to match the spectral resolution of $^{12}$CO (J = 1 -- 0) observations of Taurus, $\sim$ 0.27 km/s (Narayanan et al. 2008).

\begin{figure}
\includegraphics[width=1.0\columnwidth, clip]{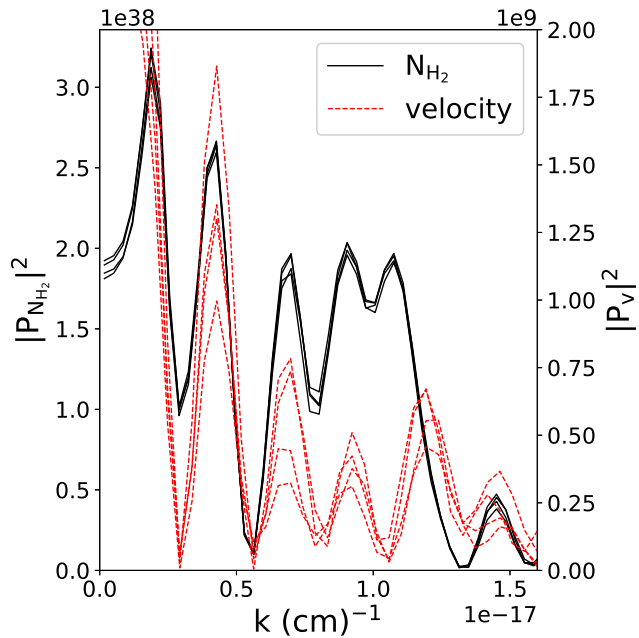}
\caption{Examples of power spectra of column density cuts perpendicular to striations (black lines) and velocity centroid power spectra (dashed red lines) from our synthetic observations at the same positions from our numerical simulations. The two power spectra peak at the same spatial frequencies, in agreement with the theoretical predictions from \S~\ref{method}.
\label{psSims}}
\end{figure}

\begin{figure}
\includegraphics[width=1.0\columnwidth, clip]{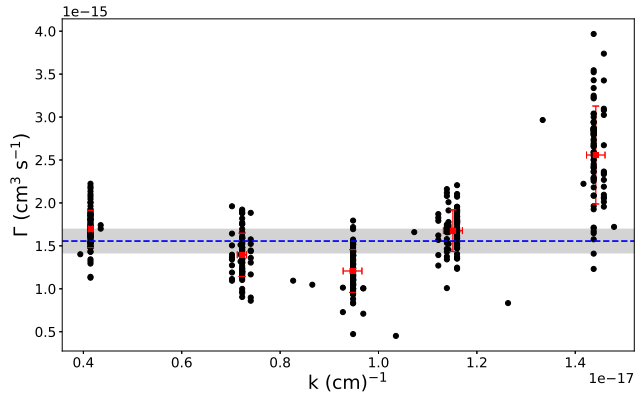}
\caption{The parameter $\Gamma$ (see Equation~\ref{definition}) as a function of the wavenumber for all peaks in the power spectra shown in Figure~\ref{psSims}. As expected from \S~\ref{method} the value of the parameter $\Gamma$ remains constant to within a factor of $\lesssim 2$ for different wavenumbers. Red square points and the errorbars are the mean and standard deviation of the black points comprising each peak. The dashed blue line shows a fit to the peaked-binned data considering a constant relation and the shaded region shows the 1$\sigma$ error of the fit.
\label{FigMoneySims}}
\end{figure}

\begin{figure}
\includegraphics[width=1.0\columnwidth, clip]{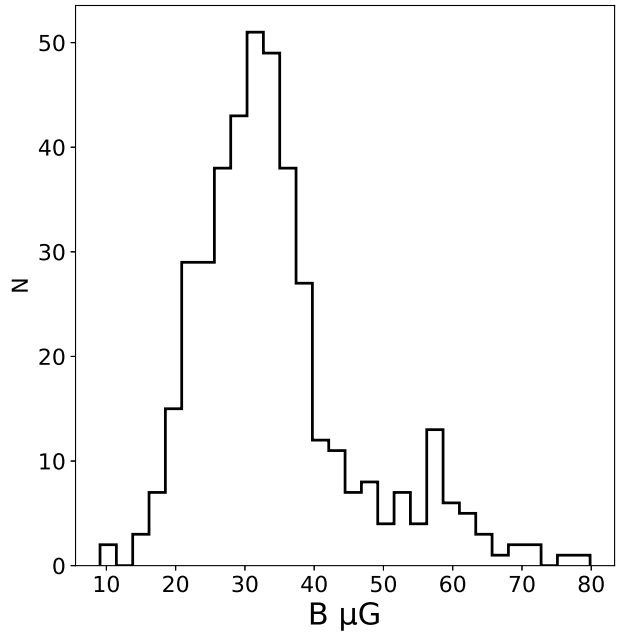}
\caption{Black histogram: distribution of magnetic fields values derived from our synthetic observations of the simulated striations (black points in Figure~\ref{FigMoneySims}).
\label{HistSims}}
\end{figure}

From the synthetic PPV cube we construct the first moment map by fitting from 1--3 Gaussian distributions to the spectral lines and computing the velocity considering a weighted average. We verified that fitting a Lorentzian or a Voigt profile does not significantly affect our results. We then compute the power spectra of cuts perpendicular to the striations in both the column density map and the first-moment map. Our results are shown in Figure~\ref{psSims}. To improve the SNR, we average the values across striations for every three adjacent cuts along the direction of the magnetic field. To avoid effects from the boundaries we considered cuts with lengths half the size of the box in the direction perpendicular to the striations. For this reason, and because of the noise added to the data, the smallest spatial frequency is not recovered in the power spectrum of velocity centroids. However, from Figure~\ref{psSims}, it can be seen that the column density and velocity centroid power spectra follow each other very well, in agreement with the theoretical expectations from \S~\ref{method}. Furthermore, the wavenumbers originally introduced in the simulation are recovered with very good precision.

We then compute the parameter $\Gamma$ (defined in Equation~\ref{definition}) by identifying the peaks in both the velocity-centroid and column-density power spectra and dividing the powers. We plot the parameter $\Gamma$ as a function of the wavenumber in Figure~\ref{FigMoneySims}. The wavenumber of each of the black points is computed as the mean of the wavenumbers of the velocity and column density power spectra at the same peak. We also compute the mean $\Gamma$ and standard deviation of the black points comprising each peak in the power spectrum and over-plotted them as red points with 1$\upsigma$ error bars. We then fit a constant relation to the red points using the orthogonal distance regression method which takes into account errors in both $x$ and $y$ coordinates (dashed blue line in Figure~\ref{FigMoneySims}). In agreement with the analysis in \S~\ref{method}, the parameter $\Gamma$ derived from the synthetic observations is equal to \( \displaystyle \frac{v_A}{N^0_{H_2}}\) and its value remains approximately constant for all peaks in the power spectrum. Deviations from the constant relation mainly arise because of the noise added to the data and the fact that the endpoints of the cuts perpendicular to the striations are not symmetric. 

The value of the magnetic field derived from the fit shown in Figure~\ref{FigMoneySims} and using Equation~\ref{mainEq} is $31\pm3 \ \rm{\upmu G}$. Thus, the true value of the magnetic field is recovered from the synthetic observations, and the method is not significantly affected by chemical or radiative transfer effects. In Figure~\ref{HistSims} we show the distribution of magnetic field values from all points in Figure~\ref{FigMoneySims}. The mode, along with the 16th and 84th percentiles, is $31^{+13}_{-7}$ $\rm{\upmu G}$, successfully recovering our input magnetic field value. 

Following the same procedure we derive the magnetic field values for the other two dynamical simulations with initial magnetic field values 10 and $100 \ \rm{\upmu G}$, respectively. The derived magnetic field value for the $10 \ \rm{\upmu G}$ simulation is $9\pm1 \ \rm{\upmu G}$ (mode and 16th and 84th percentiles are $9^{+6}_{-3} \ \rm{\upmu G}$) and for our $100 \ \rm{\upmu G}$ simulation we find a magnetic field strength of $98\pm12 \ \rm{\upmu G}$ (mode and 16th and 84th percentiles are $112^{+55}_{-42} \ \rm{\upmu G}$). 

Finally, we create a column density and a PPV cube for our reference run, seen at an angle of 45$^\circ$. The derived magnetic value for these synthetic observations is $23\pm2 \ \rm{\upmu G}$ (mode and 16th and 84th percentiles are $19^{+17}_{-8} \ \rm{\upmu G}$). This value compares very well to $19 \ \rm{\upmu G}$ which would be the value of the magnetic field directly from the simulations seen at an angle of $\uptheta = 45^{\circ}$. In Appendix~\ref{convergence} we present details from our parameter space tests, projection angle test, and results from our numerical resolution study.

\begin{figure}
\includegraphics[width=1.0\columnwidth, clip]{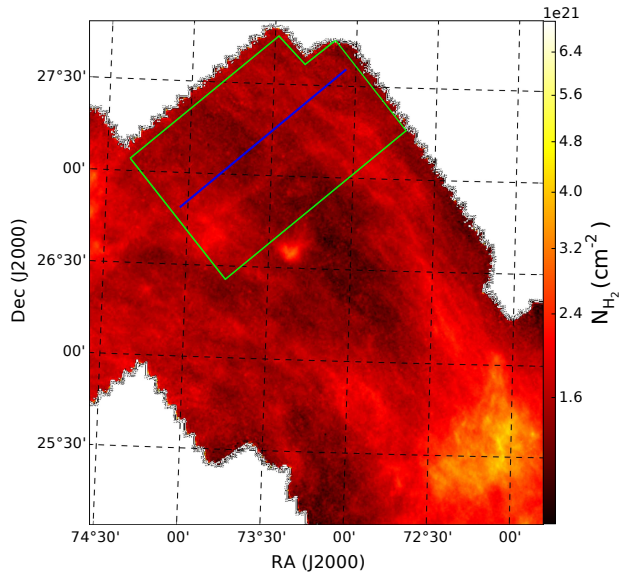}
\caption{Column density map of the striations region in Taurus where we applied our new method for deriving magnetic field values. The map was produced using $Herschel$ dust emission observations and a standard SED fitting procedure. The green polygon marks the region where we performed our analysis and the blue line shows an example of a cut perpendicular to the striations, the power spectrum of which is shown in Figure~\ref{PSobs}.
\label{striationsCD}}
\end{figure}

\begin{figure}
\includegraphics[width=1.0\columnwidth, clip]{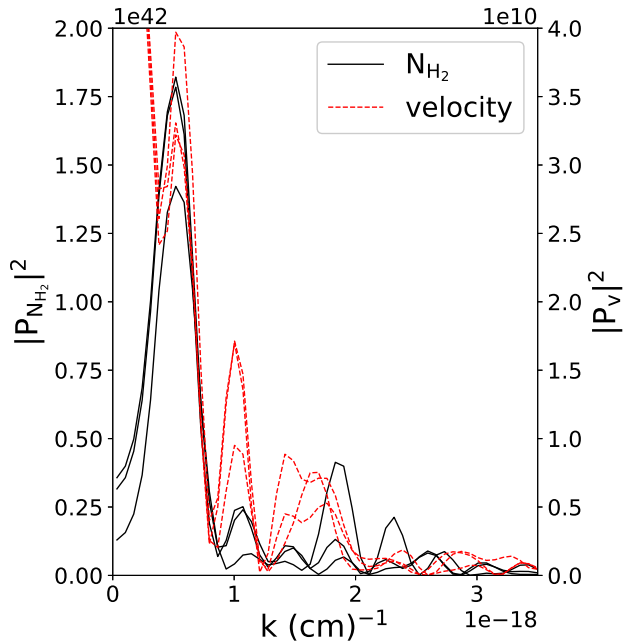}
\caption{Same as in Figure~\ref{psSims} but for the observations of the striations region in the Taurus molecular cloud. Similar to the theoretical predictions and simulations the power spectra peak at roughly the same wavenumbers.
\label{PSobs}}
\end{figure}

\section{Observations}\label{observations}

\subsection{Data}

We produce the column density and dust temperature maps of the region of striations in Taurus using the SPIRE 250, 350, and 500 $\upmu$m data from {\sl Herschel} (Griffin et al. 2010). We employed publicly available level 3 data products from the {\sl Herschel} archive. The entire Taurus region was observed within the Gould Belt keyprogram (Andre et al. 2010) as a combination of more than 10 individual tiles. Here, we only use a small cutout of the whole map, shown in Figure~\ref{striationsCD}. All dust emission maps were taken in `Parallel Mode' at a sampling rate of 10~Hz and at high speed (60$''$/s). The data provided in the archive were reduced using the pipeline of HIPE13 (Pearson et al. 2014). Each map is first processed individually in the nominal and orthogonal direction, and then combined in a second step during the map-making process. Final map reconstruction was done using the 'naive' mapmaking algorithm at the same time as the destriper (including a median correction and in bright source mode). The final gridding of the maps is 6$''$, 10$''$, 14$''$ for 250, 350, 500 $\upmu$m, respectively.  For an absolute calibration of the SPIRE maps, the Planck-HFI observations were used for the HIPE-internal zeroPointCorrection task that calculates the absolute offset for a SPIRE map, based on cross-calibration with HFI-545 and HFI-857 maps, including colour-correcting HFI to SPIRE wavebands assuming a grey-body function with fixed beta. The offsets are computed on extdPxW maps, calibrated for extended emission, with extended gain correction applied and in units of MJy sr$^{-1}$.

Column density and temperature maps were then produced at 36$''$ (all maps were smoothed to this lowest resolution of the 500 $\mu$m map) with a pixel-by-pixel SED fit from 250 $\upmu$m to 500 $\upmu$m.  We did not include the 160 $\upmu$m data so that we focus on the cold gas phase. For the SED fit, we used a specific dust opacity per unit mass (dust+gas) approximated by the power law $\kappa_\lambda \, = \,0.1 \, (\lambda/300 \,{\rm \upmu m})^\beta$ cm$^{2}$g$^{-1}$
with $\beta$=2, and left the dust temperature and column density as free parameters. We checked the SED fit of the pixels and determined from the fitted surface density the H$_2$ column density. The fit assumes a constant temperature for each pixel along the line of sight, an assumption that is not fulfilled in regions with strong temperature gradients. For the striations region in Taurus, however, we do not expect very strong temperature gradients. A detailed study of the dust properties in Orion A (Roy et al. 2013), showed that even in this brighter region, the single temperature model provides a
reasonable fit, and that the dust opacity varies with column density by up to a factor of 2. Thus, we estimate that the final uncertainties of the column density map are 20\%--30\%.

For deriving the power spectra of the velocity centroids we use $^{12}$CO (J = 1 -- 0) line emission data of the Taurus molecular cloud (Goldsmith et al. 2008). The data were obtained as part of the Five Colleges Radio Astronomy Observatory (FCRAO) survey of the Taurus molecular cloud (Narayanan et al. 2008). The spectral resolution is $\delta v$= 0.266 km/s and FCRAO's angular resolution (45'' at 115 GHz) at the distance of the Taurus cloud (127 pc) (Schlafly et al. 2014) results in a spatial resolution of $\sim$ 0.012 pc.

\begin{figure}
\includegraphics[width=1.0\columnwidth, clip]{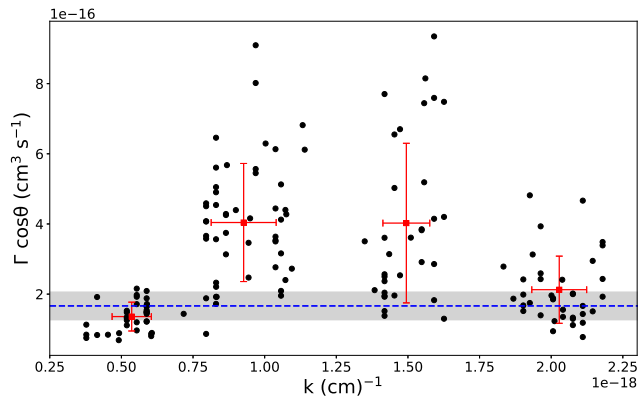}
\caption{Same as in Figure~\ref{FigMoneySims} but for the actual observations of the striations region in Taurus molecular cloud. The parameter $\Gamma$ is constant to within a factor of $\sim$ 3. A $\rm{cos\uptheta}$ is added in the y-axis label since the projection angle is unknown. 
\label{MoneyPlot}}
\end{figure}   

\subsection{Application of the new magnetic field estimation method to observations of Taurus molecular cloud}

\subsubsection{Deriving magnetic field values in Taurus}

For performing our analysis we convolved the $Herschel$ column density map and the FCRAO $^{12}$CO (J = 1 -- 0) map to the same spatial resolution (0.012 pc) using a Gaussian kernel. Spectra are smoothed to about twice the original velocity resolution ($\sim$ 0.5 km/s) in order to improve the signal-to-noise ratio. Velocity centroids are computed by fitting multiple Gaussian (up to 3) profiles to spectral lines as in the simulations. 

Examples of power spectra computed from both the column density map and velocity centroids at the same coordinates are shown in Figure~\ref{PSobs} (black and red lines respectively). As in the simulations, we have averaged every three adjacent cuts along the direction of the magnetic field in order to improve the SNR. As theoretically expected from Equations~\ref{vel} and~\ref{cd}, the two sets of power spectra have peaks at approximately the same spatial frequencies. 

\begin{figure}
\includegraphics[width=1.0\columnwidth, clip]{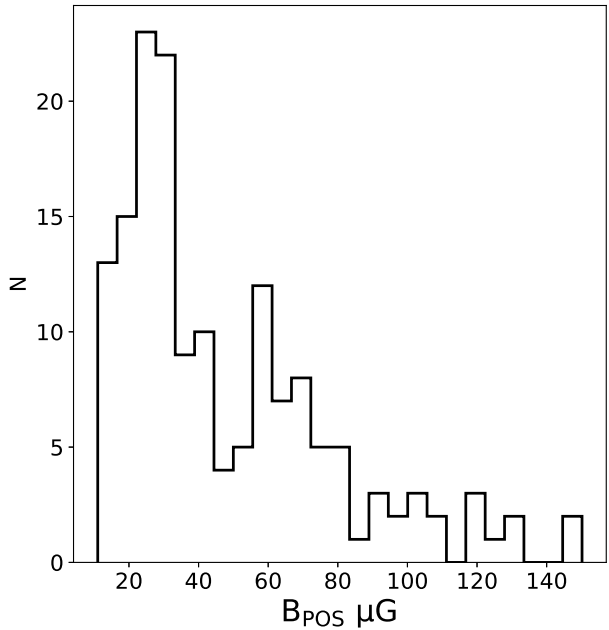}
\caption{Distribution of values for the POS magnetic field in Taurus as these are computed from all points shown in Figure~\ref{MoneyPlot}.
\label{histBs}}
\end{figure}

We compute the parameter $\Gamma$ for the first four peaks of each set of power spectra and plot our results in Figure~\ref{MoneyPlot}. Points and lines are the same as in Figure~\ref{FigMoneySims}. In essence, Figure~\ref{MoneyPlot} represents the dispersion relation of fast magnetosonic waves from observations of striations in Taurus. From the value of the intercept, assuming a mean number density of 210 $\pm ~10$ $\rm{cm^{-3}}$ (Chapman et al. 2011) and based on Equation~\ref{mainEq} we find that $B_0\rm{cos\uptheta = 27 \pm 7} ~\rm{\mu G}$. The error in $B_0\rm{cos\uptheta}$ is computed from the error of the fit and the uncertainty in density reported in Chapman et al. 2011, through error propagation. In Figure~\ref{histBs} we show the distribution of $B_0\rm{cos\uptheta}$ values as these are computed from all points. The shape of the distribution is in good agreement with the shape of the distribution produced from the synthetic observations (see also Appendix~\ref{convergence}). The mode of the distribution along with the 16th and 84th percentile is $27^{+50}_{-6}~\rm{\upmu G}$.

Using the DCF method and for approximately the same region where we performed our analysis, Chapman et al. (2011) reported a value of $\rm{12 \pm 1} ~\rm{\upmu G}$ for the POS magnetic field. Using the same polarization data but applying the Hildebrand et al. (2009) method instead of the DCF method, Chapman et al. (2011) reported a different value of $\rm{37 \pm 6} ~\rm{\upmu G}$ for the POS magnetic field. Both these estimates may be affected from systematic errors in the methods and as a result, the uncertainties in the values reported by Chapman et al. (2011) may be underestimated. Unlike the DCF method, the Hildebrand et al. (2009) method takes into account large-scale variations of the magnetic field and non-turbulent motions. Thus, regardless of the errors, the magnetic field value estimated here ($27 \pm 7 ~\rm{\upmu G}$) should be compared to the value from Chapman et al. (2011) derived using the Hildebrand et al. (2009) (method $\rm{37 \pm 6} ~\rm{\upmu G}$). These two values are in good agreement.  

\subsubsection{Deriving the ratio of the fluctuating to mean magnetic field}

Finally, we measure a ratio of the fluctuating to mean component of the magnetic field from the respective ratio of column densities (Equation~\ref{ratio}). For each column density profile, we subtract its mean value and filter the data with a low-pass filter to minimize the effects from noise. We then compute the integral of each squared filtered profile normalized over its length. Finally, we take the square root and divide by the mean value of the original profile, thus obtaining a dimensionless quantity. We find that the mean value of the fluctuating to ordered components of the magnetic field is of the order of $\sim$ 10\%. Similar values for the fluctuating to ordered component of the magnetic field have been found in the ``Brick cloud" in the Central Molecular Zone, which is also a strongly magnetized cloud (Federrath et al. 2016b). Based on turbulent molecular-cloud simulations with a strong guide field (as applicable to these types of clouds), Federrath (2016c) derived a relation between the fluctuating and ordered magnetic field components and the \Alf~Mach number. Using this relation, we find an \Alf~Mach number of 0.4--0.5 for this region. This result is in agreement with previous studies of the region with striations in Taurus (Heyer et al. 2008; Heyer \& Brunt 2012). The \Alf~Mach number is used here as a measure of the ratio of the kinetic energy to magnetic energy rather than a measure of turbulent properties.

\subsubsection{Deriving timescales of striations in Taurus}

From the derived value of the magnetic field, the value of the number density reported by Chapman et al. (2011) ($\rm{200~cm^{-3}}$) and the values of the wavenumbers, we compute the frequencies of each of the wavemodes found in the power spectrum. We find that the periods are 1.46, 0.84, 0.49 and 0.38 Myr for the first, second, third and fourth wavemode, respectively. Since the true, non-projected magnetic field value can be higher than the value derived here, the derived periods should be interpreted as upper limits. These measurements of the time-scales of striations in Taurus suggest that the largest to smallest scales of striations re-configure on time-scales of 0.73 to 0.19 Myr, that is when the phase difference with respect to present day is $\pi/2$. Making the simplification that no higher or smaller modes other than the ones analysed here are present, that particular region will be observed again as it is observed at present day in a timescale that is the least common multiple of 1.46, 0.84, 0.49 and 0.38 Myr. Each striation is not formed as a result of a single wavemode. Instead, it is formed by the interaction of all wavemodes in the system as their sum at a particular point in space and time. The structure of a region with striations at any time will depend on the powers, the wavenumbers, the phases and the period of all wavemodes, as well as the total number of wavemodes.

\section{Discussion}\label{discuss}

Deriving Equation~\ref{mainEq} we made several assumptions. Here, we discuss these assumptions and the simplifications made. As a first-order approximation, we ignored fast magnetosonic waves propagating in directions other than exactly perpendicular to the magnetic field. Waves propagating in other directions are also present in regions of striations and that intrinsically creates a non-zero spread in the orientation angle of striations with respect to the magnetic field. Panopoulou et al. 2016 found that 70\% of striations have orientations within 30$^\circ$ of the polarization segments. Malinen et al. (2016) also found strong alignment between the magnetic field and striations and a spread of $\sim$ 20$^\circ$ in their distribution. Complete distributions of the orientation angle of striations with respect to the magnetic field as probed by polarization measurements are shown in Panopoulou et al. (2016), Cox et al. (2016) and Malinen et al. (2016) for striations in Polaris, Musca and L1642 respectively. In our analysis, we only consider cuts perpendicular (or as close to perpendicular as possible) to the striations. Thus, this assumption is valid considering the manner in which we analyse striations.

A second assumption was that the sound speed squared is much smaller than the \Alf~speed squared and can thus be ignored. Generally, this is a valid assumption for all stages of molecular cloud evolution and especially for striations. Considering an upper limit of 20 K for the temperature in striations the sound speed square is $\sim$ $\rm{0.07~ (km/s)^2}$. Considering a lower limit of $\rm{10 ~\upmu G}$ for the value of magnetic field and an upper limit of $\rm{400 ~cm^{-3}}$ for the number density the \Alf~speed square is $\sim$ $\rm{0.5~ (km/s)^2}$. Thus, even considering values well outside the observational limits, this assumption holds. 

The third assumption we made was that the velocity along the direction of the magnetic field is much smaller than the velocities in the perpendicular direction. This is to be expected when fast magnetosonic waves are present since the restoring forces from the magnetic pressure will result in velocities mostly perpendicular to the magnetic field. Even if flows along the magnetic field are present, as for example in the case of accretion onto a denser structure, the velocity of these flows is \textit{not} expected to vary significantly in a cut across striations. Thus, no peaks will appear in the power spectrum because of that velocity component or peaks may appear but with very small power. Such peaks are not considered in our analysis. 

These theoretical arguments are in agreement with observations of striations. Heyer et al. (2008) performed a modified version of principle component analysis to the spectroscopic data from Narayanan et al.(2008) and Goldsmith et al.(2008). They found that there is small to no variation of velocities along the long axis of the striations while across striations the variation of velocities is strong. Thus, our third assumption that the velocity along magnetic field lines is much smaller than the velocities in the perpendicular direction is valid. 

Uncertainties in determining the POS component of the magnetic field in our new method mainly originate from three factors. The first uncertainty arises from the column density map. Assuming a constant LOS temperature for each pixel in the map and the uncertainties of beta and dust/mass ratio can lead to an uncertainty of up to 20\%--30\% in the column density map. The second uncertainty arises from the line emission data. Observations with higher spectral and spatial resolutions and SNR would greatly improve the results. However, even with the low spectral and spatial resolution and SNR line emission data, the derived value of the magnetic field is in agreement with previous estimates in Taurus. The last factor that can introduce errors is the method itself. When the endpoints of the cuts in velocity and column density cuts are highly anti-symmetric (e.g. one endpoint at +1 km/s and one at -1 km/s), the power spectrum may include spurious oscillations and/or the power in each spatial frequency may not be correctly retrieved. Thus, we advice that the cuts in both column density and in velocity are selected such that the endpoints are as symmetric as possible. Uncertainties in the distance of the cloud do not affect our derived value for the magnetic field but do slightly affect our results for the period of each wavemode. In summary, the typical  uncertainties in the magnetic field derived with our new method are of the order of 30\%.

The discovery of normal modes in the Musca molecular cloud (Tritsis \& Tassis 2018) was the first prediction of the hydromagnetic model by Tritsis \& Tassis (2016) being confirmed. In the current paper, during the analysis of the observations we were also able to retrieve the dispersion relation of fast magnetosonic waves (Figure~\ref{MoneyPlot}) and demonstrate that velocity and column density power spectra peak at the same wavenumbers. These results are a confirmation of a second prediction of the hydromagnetic model developed by Tritsis \& Tassis (2016). 

\section{Summary}\label{sum}

Robust measurements of the total magnetic field strength are of critical importance in order to set constraints on models of filament formation and cloud evolution, and are crucial in order to distinguish between different star formation theories.

Here, we used the theory of MHD waves and presented a new method to derive the POS component of the magnetic field. The method applies for regions were striations are observed. Additionally, our method enables us to measure, for the first time, time-scales in molecular clouds. Finally, we derived a new relation for measuring the fluctuating and ordered components of the magnetic field from the respective ratio of column densities. 

We derive the POS component of the magnetic field in the region of striations in Taurus. We find a value of $\rm{27 \pm 7} ~\rm{\upmu G}$. This value is in agreement with previous estimates with the Hildebrand et al. (2009) method (Chapman et al. 2011). The ratio of fluctuating to ordered component of the magnetic field was found to be $\sim$ 10\%. We find that the periods of four wavemodes identified in the power spectra are 1.46, 0.84, 0.49 and 0.38 Myrs.

\section*{Acknowledgements}

We thank G. Panopoulou for useful suggestions and discussions. We also thank the anonymous referee for comments, which significantly helped to improve this work. A.~T~and C.~F.~acknowledge funding provided by the Australian Research Council's Discovery Projects (grants~DP150104329 and~DP170100603), the ANU Futures Scheme, and the Australia-Germany Joint Research Cooperation Scheme (UA-DAAD). N.S. acknowledges support by the french ANR and the German DFG through the project "GENESIS" (ANR-16-CE92-0035-01/DFG1591/2-1). K.T. acknowledges funding from the European Research Council (ERC) under the European Union's Horizon 2020 research and innovation programme under grant agreement No 771282. The simulations and data analysis presented in this work used high-performance computing resources provided by the Leibniz Rechenzentrum and the Gauss Centre for Supercomputing (grants~pr32lo, pr48pi and GCS Large-scale project~10391), the Partnership for Advanced Computing in Europe (PRACE grant pr89mu), the Australian National Computational Infrastructure (grant~ek9), and the Pawsey Supercomputing Centre with funding from the Australian Government and the Government of Western Australia, in the framework of the National Computational Merit Allocation Scheme and the ANU Allocation Scheme. We acknowledge usage of the Metropolis HPC Facility at the CCQCN of the University of Crete, supported  by  the European Union Seventh Framework Programme (FP7-REGPOT-2012-2013-1) under grant agreement no. 316165. The $\textsc{FLASH}$ code used in this work was in part developed by the DOE NNSA-ASC OASCR Flash Center at the University of Chicago. For post processing our results we partly use the $\textsc{yt}$ analysis toolkit (Turk et al. 2011).

\appendix
\section{Parameter space and convergence tests}\label{convergence}

\begin{figure}
\includegraphics[width=1.0\columnwidth, clip]{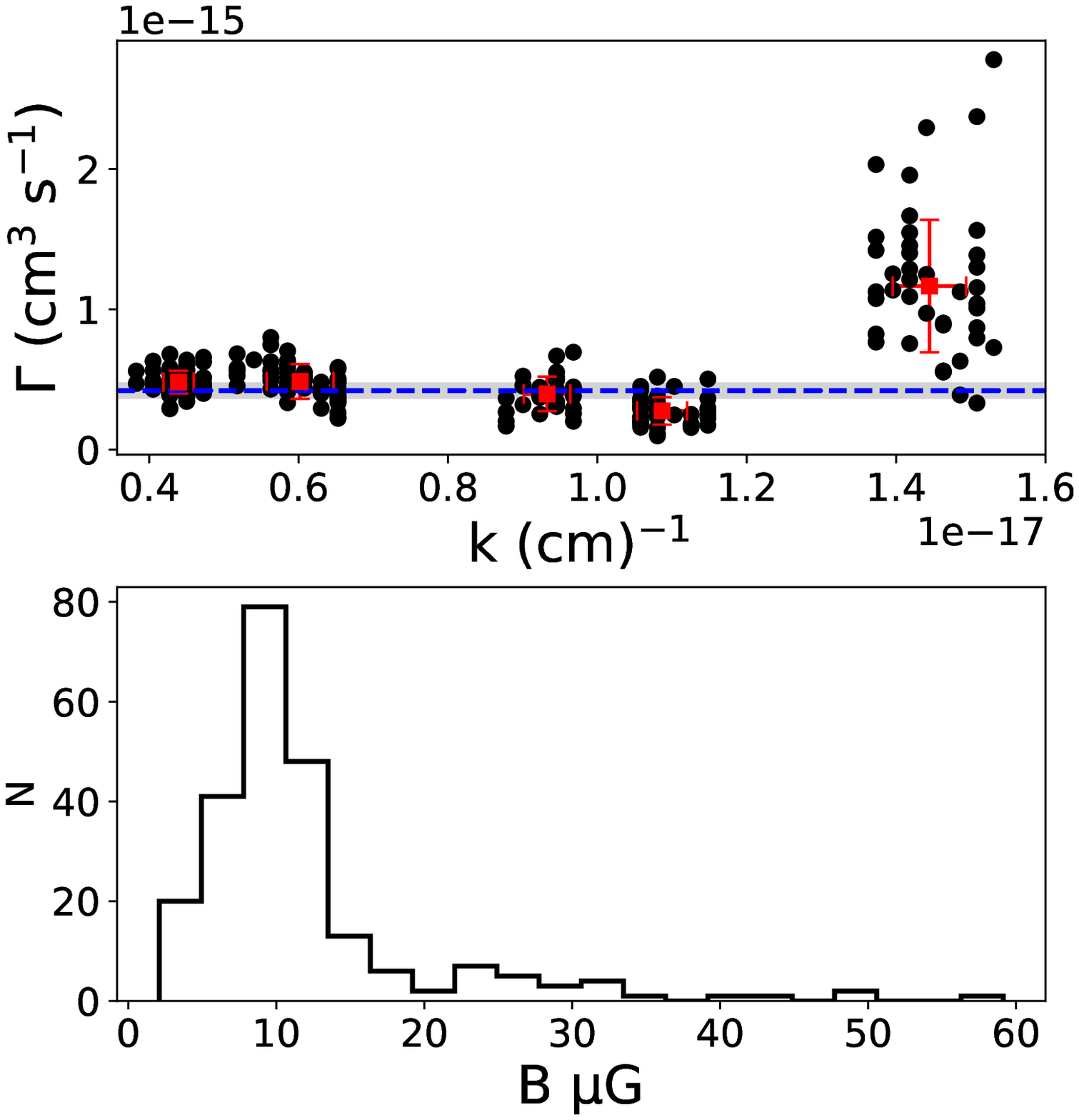}
\caption{The upper panel shows the parameter $\Gamma$ as a function of the wavenumber for our simulation with initial magnetic field value $\rm{100\ \upmu G}$. The black points have been computed from power spectra as described in the main text. The points marked with red squares and the errorbars are the mean and standard deviation of each stack of black points. The blue dashed line and the blue shaded region show the fit and error of the fit to the red points. The bottom panel shows the distribution of magnetic field values derived from the black points.
\label{ApendixB10}}
\end{figure}

\begin{figure}
\includegraphics[width=1.0\columnwidth, clip]{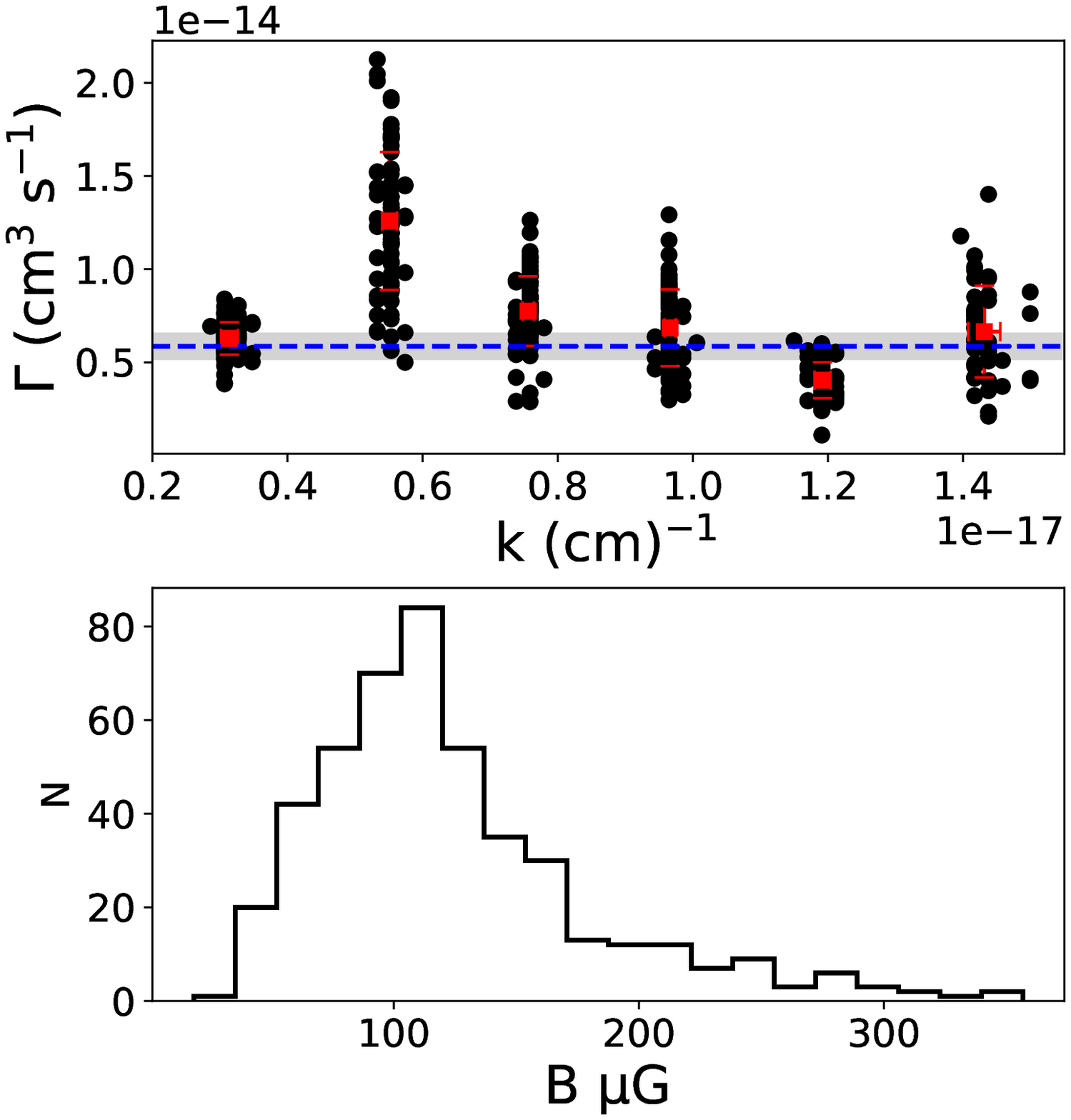}
\caption{Same as in Figure~\ref{ApendixB10} but for our simulation with initial magnetic field value $\rm{100\ \upmu G}$.
\label{ApendixB100}}
\end{figure}

In Figures~\ref{ApendixB10} and ~\ref{ApendixB100} we show the results from our simulations with initial magnetic field values 10 and 100 $\rm{\upmu G}$, respectively. In the upper panels of Figures~\ref{ApendixB10} and ~\ref{ApendixB100} we plot the parameter $\Gamma$ as a function of the wavenumber as in our reference run. In agreement to the observations and the theoretical predictions the value of the parameter $\Gamma$ remains constant regardless of the value of the magnetic field. In the lower panels, we show a histogram of magnetic field values as these are computed directly from the black points in the upper panel. The true value of the magnetic field is recovered in both cases. 

\begin{figure}
\includegraphics[width=1.0\columnwidth, clip]{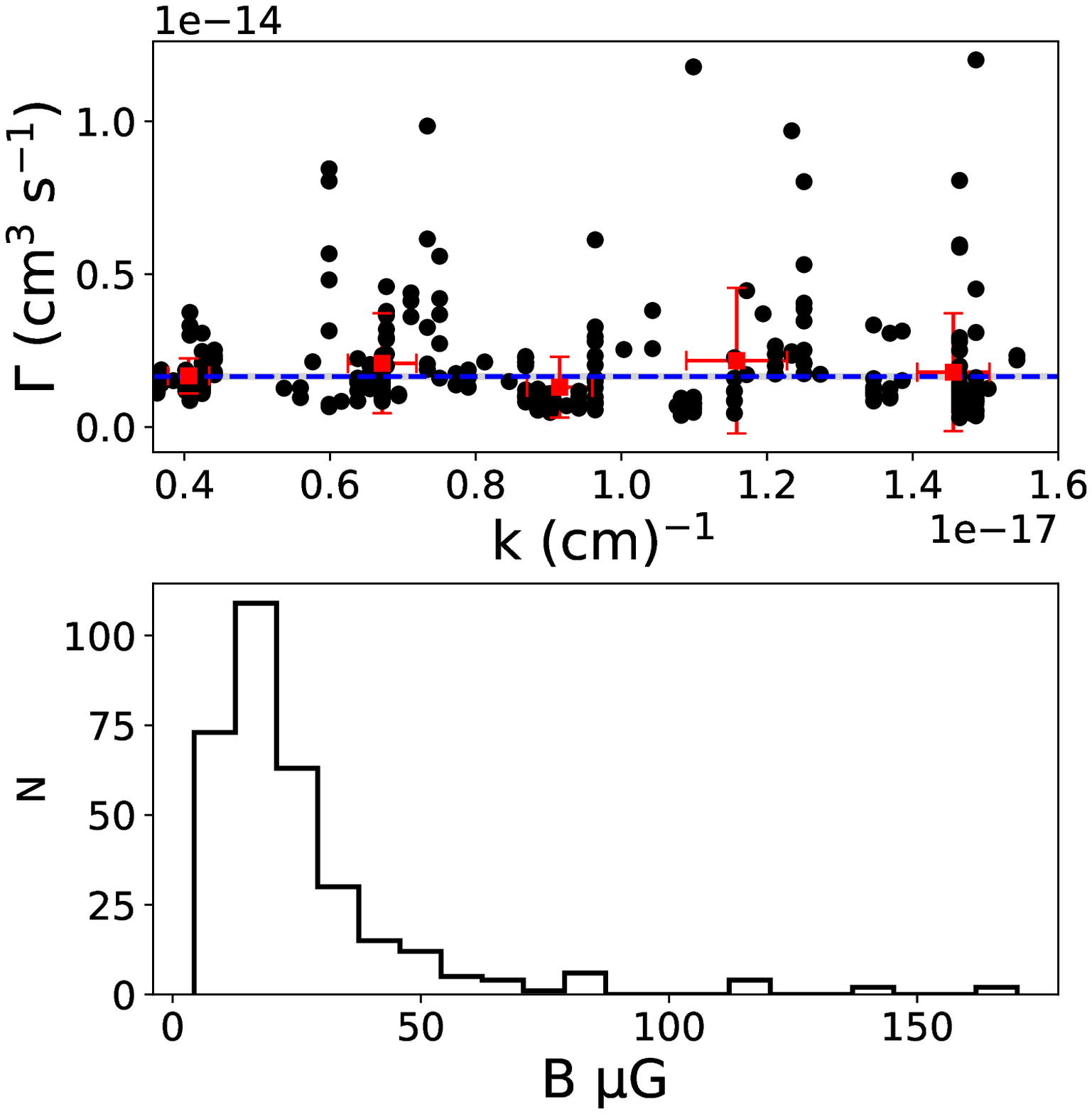}
\caption{Same as in Figure~\ref{ApendixB10} but for our simulation with initial magnetic field value $\rm{30\ \upmu G}$ when observed at an angle of 45$^\circ$.
\label{ApendixAngle45}}
\end{figure}

\begin{figure}
\includegraphics[width=1.0\columnwidth, clip]{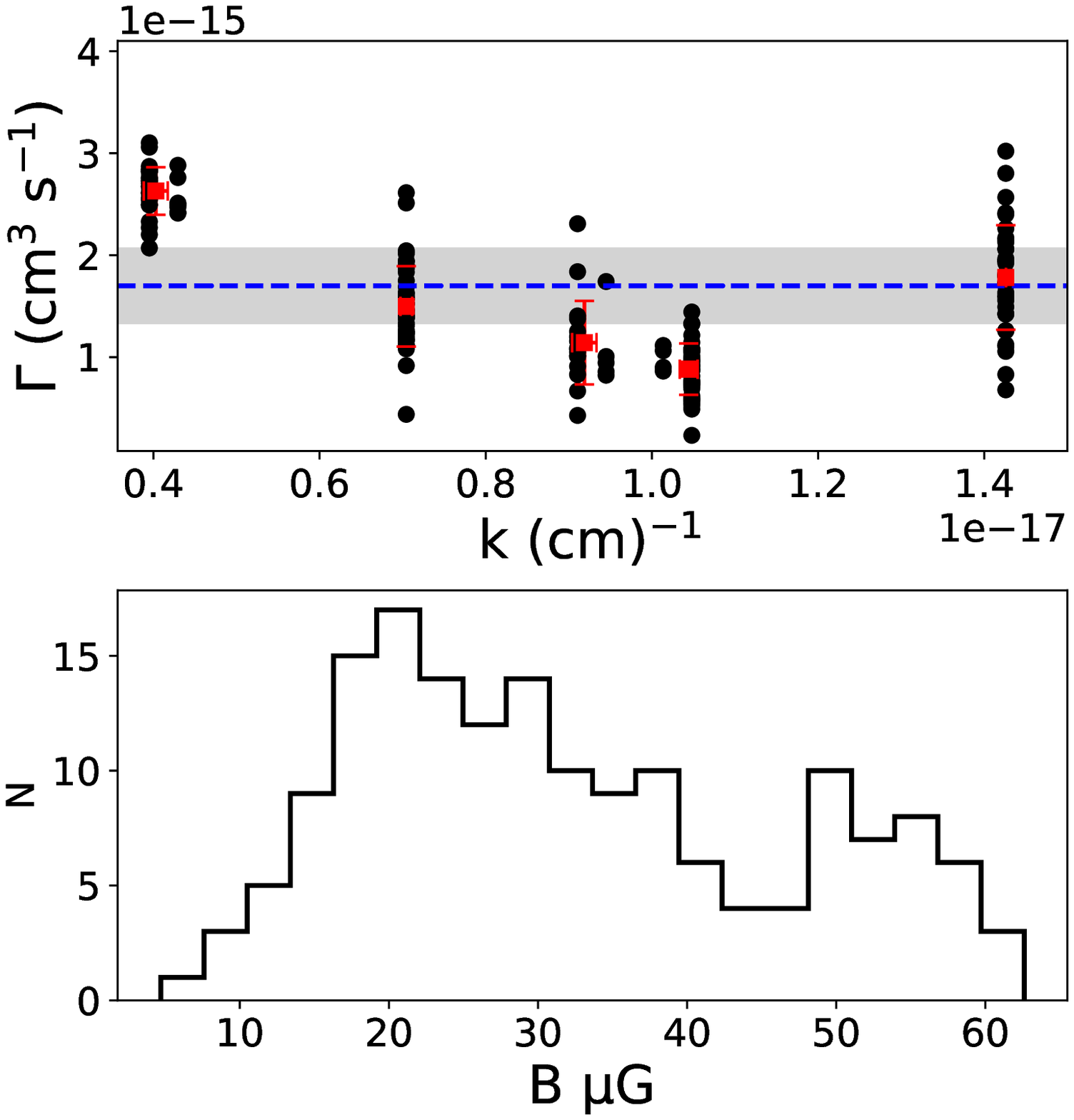}
\caption{Same as in Figure~\ref{ApendixB10} but for our simulation with initial magnetic field value $\rm{30\ \upmu G}$ and with half the resolution of our reference run.
\label{Rconvergence}}
\end{figure}

In Figure~\ref{ApendixAngle45} we show our results when the cloud is observed such that the angle between the magnetic field and the POS is $\uptheta$ = 45$^\circ$. The value of the parameter $\Gamma$ is smaller by a factor of $\rm{cos}\uptheta = $ 0.7 with respect to the case when the projection angle is 0, again in agreement with the theoretical predictions. Finally, we have repeated our analysis for our simulation with half the original resolution (128 $\times$ 128 grid points) of our reference run. The derived magnetic field value is $34\pm8$ $\rm{\upmu G}$, in agreement with the higher resolution simulation. Our results are shown in Figure~\ref{Rconvergence}.

\end{document}